\documentclass[%
 reprint,
superscriptaddress,
 amsmath,amssymb,
 aps,
pra,
]{revtex4-2}

\usepackage{graphicx}
\usepackage{dcolumn}
\usepackage{bm}

\newcommand{\bnen}{\begin{equation}}
\newcommand{\eden}{\end{equation}}
\newcommand{\bean}{\begin{eqnarray}}
\newcommand{\eean}{\end{eqnarray}}
\newcommand{\bnsn}{\begin{subequations}}
\newcommand{\edsn}{\end{subequations}}

\newcommand{\bea}{\begin{eqnarray*}}
\newcommand{\eea}{\end{eqnarray*}}
\newcommand{\bne}{\begin{equation*}}
\newcommand{\ede}{\end{equation*}}

\usepackage{verbatim}
\usepackage[utf8]{inputenc}
\usepackage[english]{babel}
\usepackage[T1]{fontenc}
\usepackage{amsfonts}
\usepackage{mathtools}
\usepackage{hyperref}
\usepackage{float}
\usepackage{tikz}
\usepackage{lipsum}

\usepackage{mathtools}
\usepackage{multirow}
\usepackage{amsmath}
\usepackage{physics}

\begin{document}

\preprint{APS/123-QED}

\title{Weyl points in ball-and-spring mechanical systems}

\author{Zolt\'an Guba}
\affiliation{Department of Theoretical Physics, Institute of Physics, Budapest University of Technology and Economics, Műegyetem rkp. 3., H-1111 Budapest, Hungary}
\author{Gy\"orgy Frank}
\affiliation{Department of Theoretical Physics, Institute of Physics, Budapest University of Technology and Economics, Műegyetem rkp. 3., H-1111 Budapest, Hungary}
\author{Gerg\H o Pint\'er}
\affiliation{Department of Theoretical Physics, Institute of Physics, Budapest University of Technology and Economics, Műegyetem rkp. 3., H-1111 Budapest, Hungary}
\author{Andr\'as P\'alyi}
\affiliation{Department of Theoretical Physics, Institute of Physics, Budapest University of Technology and Economics, Műegyetem rkp. 3., H-1111 Budapest, Hungary}
\affiliation{MTA-BME Quantum Dynamics and Correlations Research Group, Műegyetem rkp. 3., H-1111 Budapest, Hungary}

\date{\today}

\begin{abstract}
Degeneracy points of parameter-dependent Hermitian matrices play a fundamental role in quantum physics, as illustrated by the concept of Berry phase in quantum dynamics, the Weyl semimetals in condensed-matter physics, and the robust ground-state degeneracies in topologically ordered quantum systems. 
Here, we construct simple ball-and-spring mechanical systems, whose eigenfrequency degeneracies mimic the behaviour of degeneracy points of electronic band structures. 
These classical-mechanical arrangements can be viewed as ‘de-quantized’ versions of Weyl Josephson circuits, i.e., superconducting nanostructures proposed recently to mimic band structure effects of Weyl semimetals. 
In the mechanical setups we study, we identify degeneracy patterns beyond simple Weyl points, including the chirality flip effect and a quadratic degeneracy point.
Our theoretical work is a step toward simple and illustrative table-top experiments exploring topological and differential geometrical aspects of physics.
\end{abstract}

\maketitle

\tableofcontents

\section{
\label{sec:intro}
Introduction}

Topological semimetals have attracted significant attention in recent years due to their unique electronic properties and potential applications in next-generation electronics.
The electronic band structure of such materials exhibits degeneracy points, e.g., Weyl points, in their band structure, at which the energy eigenvalues disperse linearly. 
These points act as sources or sinks of Berry curvature, which leads to a number of interesting phenomena, including Fermi arc surface states, 
chiral anomaly, and anomalous Hall effect
 \cite{Bernevig_2018, Armitage_2018, Yan_review}.
Beyond Weyl-point physics, robust level degeneracies are essential ingredients of topological insulators \cite{Asboth2016}, topological quantum computing 
\cite{KitaevMajorana},
and topologically ordered systems as well \cite{KitaevToric, Wen_book}.

The complexity of real materials with Weyl points in their electronic band structures often hinders the observation of the associated geometrical and topological effects.
Because of that, the physical characteristics of Weyl points are often investigated using metamaterials, e.g., engineered, artificial crystals whose phononic or photonic band structures possess Weyl points \cite{Yang, Li, Chen_2022, Luo}.

Alternatively, Weyl points arise and can be studied in quantum systems with at least 3 control parameters. 
For example, multiply-connected superconducting devices \cite{Zur_rev}, including multi-terminal Josephson junctions \cite{Riwar} and the recently proposed Weyl Josephson circuits \cite{Fatemi}, can emulate Weyl semimetal band structures, where, e.g., the magnetic fluxes piercing the loops of the circuit correspond to the wave vectors of a band structure.
Though multiply connected superconductors are a promising testbed to emulate and investigate topologically non-trivial band structures, the realization of such experiments requires costly, advanced, and challenging fabrication, as well as millikelvin cooling technology. 

In this work, we show that Weyl points and much of their rich phenomenology can be realized in  mechanical ball-and-spring systems, potentially leading to much simpler and much less costly table-top experiments on Weyl point physics. 
Weyl points arise in the parameter-dependent frequency spectrum of the normal-mode oscillations of the proposed ball-and-spring systems.
In the first setup we propose (System A), we illustrate the appearance of Weyl points, their movement, their creation and annihilation, highlighting the case when the spatial symmetry of the setup governs the creation-annihilation process, enabling the \emph{chirality flip} effect \cite{Konye}. 
In the second setup we propose (System B), we show that the parameter-dependent effective dynamical matrix is analogous to the wave-vector dependent effective Hamiltonian of bilayer graphene \cite{McCann}, which exhibits a non-generic degeneracy point with topological charge of 2 and local multiplicity (birth quota) of 4 \cite{Pinter}. 

The rest of the paper is structured as follows. 
In Sec.~\ref{sec:preliminaries} we introduce preliminary concepts and highlight relevant background to make this work self-contained. 
In Sec.~\ref{sec:systemA}, we introduce a simple mechanical system (System A) whose vibrational spectrum contains Weyl points, and we demonstrate their movement and creation/annihilation, and the chirality flip effect. 
In Sec.~\ref{sec:systemB}, we discuss the appearance of a charge-2 Weyl point in another classical mechanical system (System B). 
In Sec.~\ref{sec:Discussion} we discuss the relation to prior work as well as open follow-up problems, while Sec.~\ref{sec:conc} provides our conclusions. 

\section{Preliminaries}
\label{sec:preliminaries}

\begin{figure*}
    \centering
    \centering
    \includegraphics{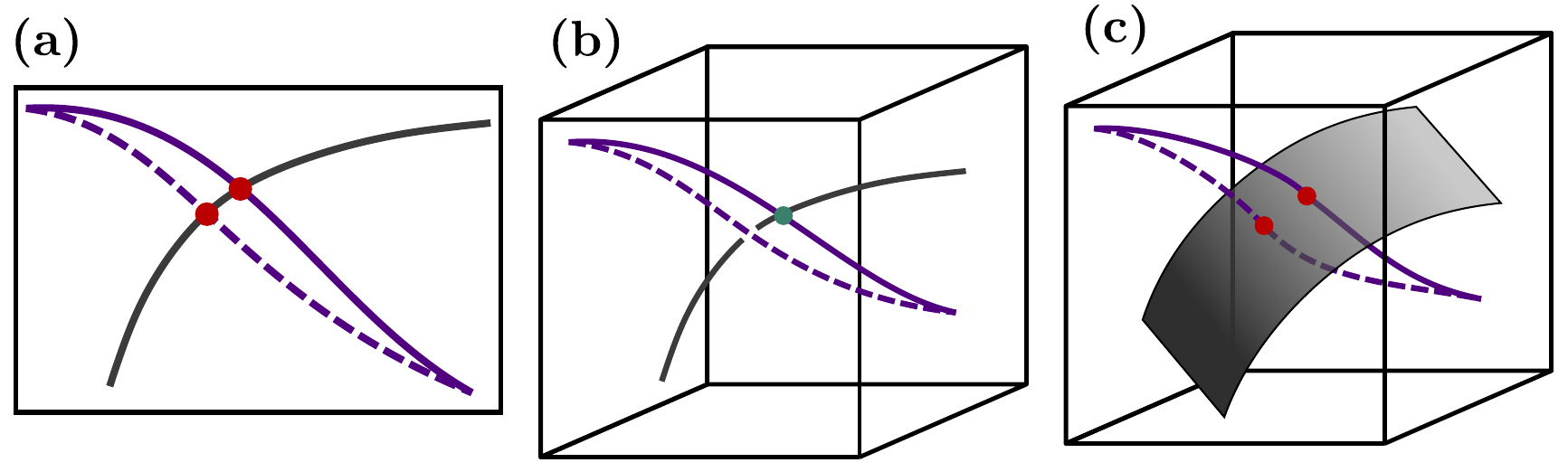}
    \caption{Protected and non-protected intersection points. \textbf{(a)} Two generic intersecting curves (solid grey and purple lines) on the Euclidean plane. 
    The intersection point is denoted with the upper red dot. 
    Upon a small deformation of the purple curve (denoted with the dashed curve), the two curves (solid grey and dashed purple) still intersect: the intersection point is `protected'.
    \textbf{(b)} Two intersecting curves (purple and grey solid lines) in 3D Euclidean space. 
    The intersection point is denoted with a green dot.
    Upon a small deformation of the purple curve (denoted with the dashed curve), the two curves (solid gray and dashed purple) avoid each other; 
    the intersection point is `not protected'. 
    \textbf{(c)} Curve and surface embedded in 3D Euclidean space, intersecting transversally.
    The intersection point is denoted with a red dot. Upon a small deformation of the purple curve (denoted with the dashed curve), the intersection point is shifted, but still exists;
    the intersection point is `protected'.
    Panels (a) and (c) show minimal models of `protected' Weyl points of Weyl semimetals: the host space models the space of Hermitian matrices, the solid purple line models the image of the Hamiltonian map from the Brillouin zone to the space of Hermitian matrices, and the black line/surface models the stratum of the matrix space with a twofold degeneracy. 
    }
    \label{fig:lines_surfaces}
\end{figure*}

To ensure that our terminology is well-defined, and to make this work self-contained, we collect a few preliminary concepts and relations in this section.
We start by connecting elementary geometry with the `topological protection' or `robustness' of Weyl points in quasiparticle band structures of 3D crystals. 

Consider two arbitrary intersecting curves on the Euclidean plane, as shown by the solid lines in Fig.~\ref{fig:lines_surfaces}a. Notice that in the vicinity of the intersection point, the two curves are well approximated by straight lines, which enclose a nonzero angle. This type of intersection of two curves on the plane is called \emph{transversal}.
In this case, transversality implies robustness, in the following sense:
If we slightly deform the solid purple curve into the dashed purple curve, then the intersection point still exists between the solid black and dashed purple curves, and these curves still enclose a nonzero angle in the vicinity of the intersection point.  
This robustness of the intersection point (and the local behavior around the intersection point) is sometimes referred to as `protection' against  small `deformations' or `perturbations'. 

Consider a slightly different situation: two transversally intersecting curves in three-dimensional (3D) Euclidean space, as shown by the solid lines in Fig.~\ref{fig:lines_surfaces}b. Such an intersection is transversal, but it is not protected against small deformations. As shown in Fig.~\ref{fig:lines_surfaces}b, a small deformation of one of the lines (dashed purple line) can lead to an avoidance of the two curves, and correspondingly, the disappearance of the intersection point. 
However, if we consider a transversal intersection point of a line and a surface embedded in 3D Euclidean space, as shown in Fig.~\ref{fig:lines_surfaces}c, then the intersection point is protected again.

These three simple examples of Fig.~\ref{fig:lines_surfaces} reveal an interesting property of an isolated transversal intersection point of two manifolds embedded in a host manifold. 
Namely, if the dimension of the host manifold $D$ equals the sum of the dimensions of the two embedded manifolds $D_1$ and $D_2$, that is, $D =  D_1 + D_2$, then the intersection point is protected against any small deformation. However, if the dimension of the host manifold is greater than the sum of the two embedded manifolds, that is, $D >  D_1 + D_2$, then the intersection point is not protected.

These observations lead to the often-stated conclusion that a Weyl point in the band structure of a three-dimensional crystal is protected against small deformations of the Hamiltonian. In fact, a Hamiltonian describing such a band structure is a map from the crystal’s Brillouin zone (essentially, a 3D torus) to the space of $n \times n$ Hermitian matrices, with an integer $n \geq 2$. The matrix space plays the role of the host manifold. A Hermitian matrix can be described by the real and imaginary parts of its matrix elements, that is, this matrix space has dimension $D = n^2$.
Within this matrix space, the matrices with a twofold eigenvalue degeneracy (i.e., the matrices with $i$th and $(i+1)$th eigenvalues being equal, but different from all other eigenvalues) form a manifold of codimension 3 \cite{NeumannWigner, ArnoldHermitian, Zur_rev}, that is, dimension $D_1 = n^2-3$.
This manifold is sometimes called a degeneracy stratum.
Furthermore, the image of the 3-dimensional Brillouin zone via the Hamiltonian map is a manifold in the matrix space, also of dimension $D_2 = 3$.
A Weyl point, i.e., a twofold degeneracy of the band structure, with linear dispersion in its vicinity, is in fact a transversal intersection point between the $n^2-3$-dimensional degeneracy stratum and the 3-dimensional image of the Brillouin Zone. 
This, together with the observation in the preceding paragraph, imply the robustness of a Weyl point against small perturbations, as stated above.

In the context of electronic (or more generally, phononic, photonic, magnonic, etc.) band structures, the Hamiltonian might depend not only on the wave vector but also on other physical parameters, such as mechanical strain applied to the crystal. In such a case, it is interesting to consider how the Weyl points move, merge or are born, as mechanical strain is varied. We will use the terminology that parameters characterising the position of the Weyl points are called \emph{configuration parameters}, and all other parameters are called \emph{control parameters}. In the above example, the configuration space (i.e., the space of configuration parameters) is the Brillouin zone, and the control space (the space of control parameters) is a six-dimensional space describing the mechanical strain tensor, which is a $3 \times 3$ symmetric real matrix. 

So far, we discussed Weyl points in the context of band structures. However, the mathematical structures used in the above arguments are more general, they apply to parameter-dependent Hermitian matrices in general. Hence, Weyl points arise not only in band structures, but more generally, e.g., in parameter-dependent quantum systems \cite{NeumannWigner, Zur_rev, Scherubl, Stenger}.
If the parameter space of the quantum system is 3-dimensional, then twofold degenerate band crossings do arise typically. If the parameter space has more than 3 dimensions, then one can identify a 3 dimensional parameter manifold as the configuration space, and the complementary parameter manifold as the control space. In this picture, the Weyl points are moving in the configurational space as the control parameters are varied.
Although for 3D band structures, the natural configuration space is the Brillouin zone, which is a 3D torus, in a more general setting, the configuration space does not have to be a torus, see, e.g., \cite{Scherubl,Stenger}.

As argued above, the appearance of Weyl points in band structures of 3D materials is rather natural. 
However, other types of twofold degeneracies can be achieved by fine-tuning or symmetries. 
For example, it has been argued in \cite{Yu_2022}
that crystalline symmetries can `stabilize' or `protect' three other types of twofold degeneracy points, which are called charge-2 Weyl point, charge-3 Weyl point, and charge-4 Weyl point.
These degeneracy points differ from Weyl points (which are sometimes called charge-1 Weyl points) in the following respects: (1) their dispersion relation is non-linear, (2) they are not robust against symmetry-breaking perturbations, i.e., they can be `dissolved' to a set of Weyl points if the Hamiltonian is perturbed such that the symmetry is not preserved; correspondingly, they are often referred to as \emph{non-protected} or \emph{non-generic}. 

The study of Weyl points and non-generic degeneracy structures have been proposed recently in multi-terminal Josephson circuits 
\cite{Riwar, Fatemi}.
In particular, in the proposal of Weyl Josephson Circuits
\cite{Fatemi}, whose quantum-mechanical Hamiltonian is a parameter-dependent Hermitian matrix, magnetic fluxes and gate voltages play the role of the  parameters. Furthermore, the corresponding parameter spaces are cyclic, similarly to the Brillouin zone of crystals. 
Because of the strong analogy, it has been argued that Weyl Josephson Circuits can emulate Weyl points and non-generic degeneracy patterns in band structures 
\cite{Fatemi,Frankteleportation}.
Such non-generic degeneracy patterns may include the creation or annihilation of Weyl points, the presence of non-generic isolated degeneracy points and their dissolution to Weyl points upon deformation of the Hamiltonian \cite{ChenFang, Pinter},
nodal lines 
\cite{Fatemi}
or surfaces, Weyl-point teleportation \cite{Frankteleportation},
symmetry-constrained chirality flip processes \cite{Konye}, etc.

Our present work builds upon the latter idea of emulating band-structure effects, but translates it to a simple  classical mechanical setting: a system of linearly coupled harmonic oscillators, or more specifically, a ball-and-spring system. Such a system is described by a dynamical matrix $D$, which is a real symmetric $n \times n$ matrix, where the integer $n \geq 2$ is the number of coordinates. For example, in the setup in Fig.~\ref{fig:systemA_figures}a, the point mass (green circle) can move in two dimensions, hence $n=2$.
Note also that the dynamical matrix $D$ has non-negative eigenvalues $\lambda_1, \dots,\lambda_n \geq 0$, whose square roots $\omega_j = \sqrt{\lambda_j}, \, (j = 1, \dots, n)$ provide the normal-mode eigenfrequencies. 

In fact, the dynamical matrix is a function of the parameters characterising the system, $D = D(\mathbf{p})$, where $\mathbf{p}$ is the vector of parameters, e.g., spring constants and unstretched spring lengths.
Notice that this setting is similar to that of parameter-dependent Hermitian matrices, with the important difference that the surface in the space of real symmetric matrices on which an eigenvalue is twofold degenerate has codimension 2 
\cite{NeumannWigner},
unlike the Hermitian case discussed above, with codimension 3.

As a consequence, 2D variants of Weyl points, i.e., point-like twofold frequency degeneracies with linear dispersion in their vicinity, typically appear in classical mechanical systems described by a dynamical matrix $D(\mathbf{p})$, when two parameters of the vector $\mathbf{p}$ are varied. 
This observation suggests that to emulate some of the band structure effects listed above, it might be sufficient to engineer tunable classical mechanical ball-and-spring systems. This is what we pursue in this work. 

\section{Mechanical 2D Weyl points, their creation and annihilation, and the chirality flip}
\label{sec:systemA}

\begin{figure*}
    \centering
    \includegraphics{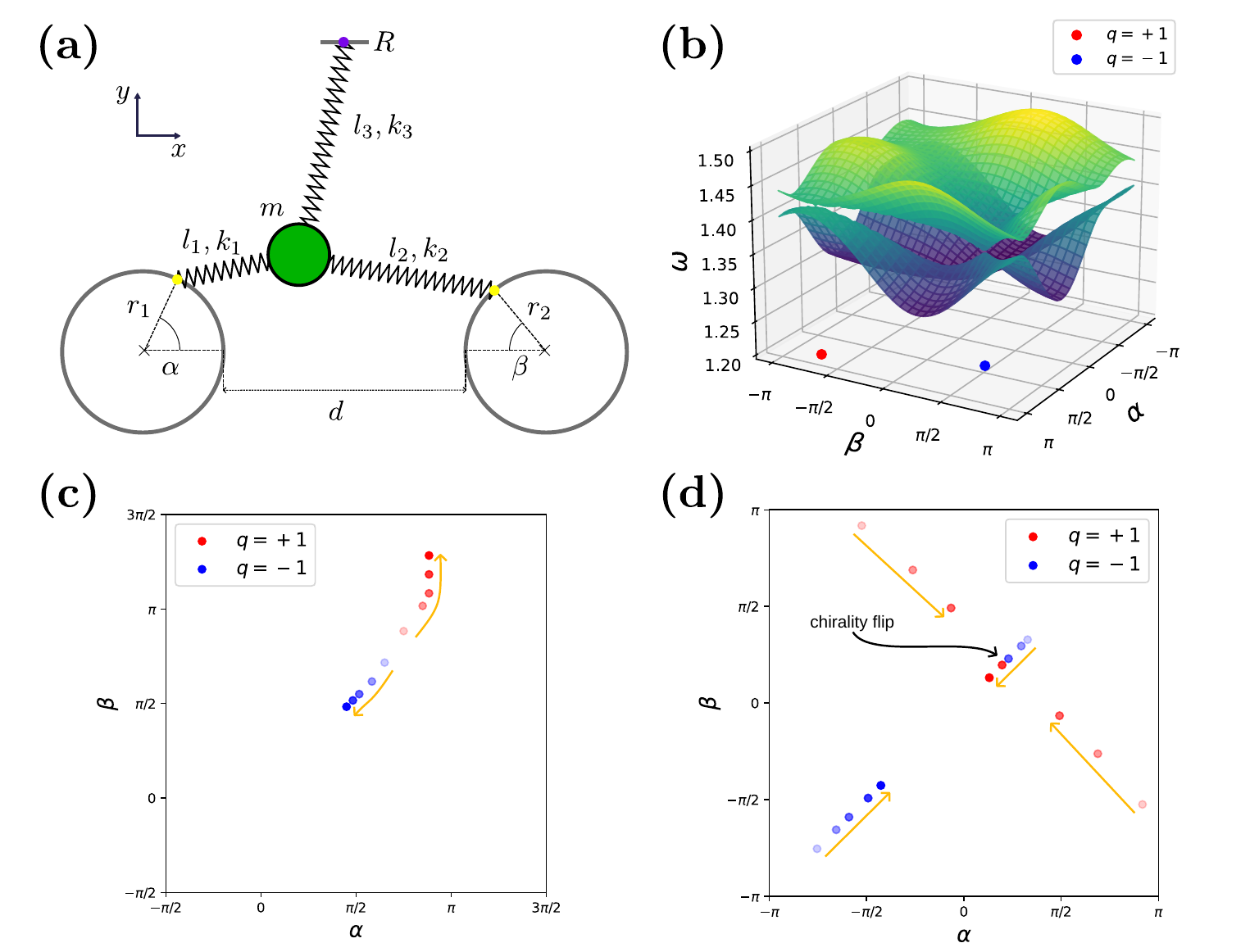}
    \caption{
    Weyl points and their movement and creation/annihilation in a mechanical system.
    \textbf{(a)} Layout of System A. Green circle: point mass $m$ moving in the  plane. 
    \textbf{(b)} Mode eigenfrequencies as function of configuration parameters $\alpha$ and $\beta$, exhibiting a positively (negatively) charged Weyl point depicted as the red (blue) point. 
    Control parameters:
    $r_1 = 2.6$, $r_2 = 2$, $k_1 = k_2 = 1$, $l_1 = l_2 = 6$, $d = 20$, $R = 10$, $l_3 = 10$ and $k_3 = 0.86$. 
    Eigenfrequencies were obtained numerically for discrete $(\alpha,\beta)$ values on a $61 \times 61$ grid. 
    \textbf{(c)} Weyl points and their movement in the configurational space  with varying spring constant $k_3 \in (0.76, 0.78, 0.80, 0.82, 0.84, 0.86)$
    (further control parameters as in Fig.~\ref{fig:systemA_figures}b).
    Darker circles depict Weyl points at higher $k_3$ values. 
    \textbf{(d)} Symmetry-induced chirality flip effect. 
    Evolution of Weyl points is shown as spring constant is changed as $k_3 \in (0.96, 1.0, 1.04, 1.08, 1.12)$, using the symmetric control-parameter set $r_1 = r_2 = 2$, $k_1 = k_2 = 1$, $l_1 = l_2 = 6$, $d = 20$, $R = 10$, $l_3 = 10$ and changing  Darker circles denote the Weyl points for higher $k_3$ values. 
    For $k_3 = 0.96$, there are 4 Weyl points in the spectrum (faintest points). 
    By increasing $k_3$, the Weyl points move in the spectrum along symmetry-constrained trajectories. 
    By changing $k_3=1.04$ to $k_3 = 1.08$ the chirality of one Weyl point flips while merging with 2 Weyl points of charge +1. 
    The Weyl points located outside the symmetry line are mirror symmetric partners of each other due to the mirror symmetry of the control parameters. }
    \label{fig:systemA_figures}
\end{figure*}

In this section, we introduce a classical system composed of balls and springs, whose vibrational spectrum emulates a number of Weyl-point-related features of electronic band structures of crystals. 
To enhance the analogy between our setup and band structures, we engineer the configurational parameter space to have torus topology, similarly to the Brillouin zone. 
Our mechanical system, depicted in Fig.~\ref{fig:systemA_figures}a, to be referred to as System A, exhibits the following features: (i) the existence of 2D Weyl points in the configurational space for fixed control-parameter vector. (ii) The movement of 2D Weyl points in the configurational space, as the control vector is varied. (iii) Creation and annihilation of oppositely charged Weyl-point pairs as the control vector is varied. (iv) the `chirality flip' effect \cite{Konye}, which is a special type of Weyl-point creation/annihilation promoted by the symmetry of the system. 

The setup, shown in Fig.~\ref{fig:systemA_figures}a, consists of a point mass $m$, three springs, and two rings.
The motion of the mass is restricted to the plane of the figure, and the orientation of the $x$-$y$ reference frame is also shown.
The centers of the rings are located at the points 
$(x,y) = (-d/2-r_1,0)$ and 
$(x,y) = (d/2+r_2,0)$, and the radii of the rings are $r_1$ and $r_2$, respectively.
On each ring, a spring is attached to a point of the ring, and the other ends of the two springs are attached to the mass.
The two suspension points on the two rings are parametrised by the angles $\alpha$ and $\beta$. 
The top end of the third spring is attached to the suspension point at $(0,R)$.
The springs are characterised by their spring constants $k_j$ and rest lengths $l_j$. 
The mass, whose vibrational modes we are interested in, is attached to these springs, and its equilibrium position depends on the system parameters.

The number of parameters of this setup is 13. In what follows we consider the angle parameters $\alpha$ and $\beta$ as configuration parameters,  and call others the control parameters. 
Therefore, the topology of the configuration space $\mathcal{B}_A = [-\pi,\pi) \times [- \pi, \pi)$ is a torus, similar to the Brillouin zone of a 2D crystal.
The 11 control parameters are positive real numbers which we collect into a vector $\mathbf{t}$.
In what follows we use SI units for all physical quantities and omit units when specifying parameter values.

The key quantities we will describe here are the eigenfrequencies of the small oscillations (normal modes) of the mass in this setup.
For a fixed set of control parameters (i.e., for a fixed control vector $\mathbf{t}$), we define the mapping $\omega_\mathbf{t}: \mathcal{B}_A \rightarrow \mathbb{R}_+^2$ 
that assigns the eigenfrequencies of the system to each point of the configuration space in such a way that the first component is the greater eigenfrequency. 
The eigenfrequencies are the square roots of the eigenvalues of the dynamical matrix of the system. 

Since System A consists of a single mass with its motion restricted to two dimensions, its  dynamical matrix is a matrix in $\mathrm{Sym}_2(\mathbb{R})$, the vector space of $2 \times 2$ real symmetric matrices.
It is instructive to decompose the dynamical matrix as a linear combination of Pauli matrices; this reads
\begin{equation}
\label{eq:pauli_decomposition}
    D(\alpha, \beta) = d_x(\alpha, \beta) \sigma_x + d_z(\alpha, \beta) \sigma_z,
\end{equation}
where $\sigma_x$ and $\sigma_z$ are the Pauli $X$ and $Z$ matrices, the dependence on $\alpha$ and $\beta$ is explicitly denoted, while the dependence on $\mathbf{t}$ is omitted for brevity. 

The normal modes of this mechanical system exhibit the Weyl-point features (i)-(iv) listed above, as shown in Figs.~\ref{fig:systemA_figures}b, c, d. 
To obtain these results, we have computed the dynamical matrix, and from that, the eigenfrequency spectrum $\omega_\mathbf{t}$, as described in Appendix~\ref{appendix:systema}. 

(i) \emph{Existence of Weyl points.}
The eigenfrequency spectrum of System A is plotted as function of the configurational parameters $\alpha$ and $\beta$, for a fixed control vector, in Fig.~\ref{fig:systemA_figures}b, see caption for parameter values.  
The spectrum contains two 2D Weyl points, indicated as the red and blue points, where the vibrational eigenfrequencies are degenerate.  
Each of the band crossing points seen in Fig.~\ref{fig:systemA_figures}b has a nonzero topological charge.
The topological charge of a 2D Weyl point, analogous to the Chern number of Weyl points in 3D band structures, is the winding number of the vector field $(d_x,d_z)$ for a loop in the configurational space enclosing the degeneracy point.
In Fig.~\ref{fig:systemA_figures}b, red (blue) points denote topological charge $+1$ ($-1$).
As illustrated in the figure, the sum of the topological charges of the Weyl points is zero. 

(ii) \emph{Movement of Weyl points.}
By changing the control parameters $\mathbf{t}$, the Weyl points trace out a trajectory in the configuration space. 
This is shown in Fig.~\ref{fig:systemA_figures}c, as the spring constant $k_3$ is varied, all other control parameters being fixed. 
The blue/red colors correspond to the topological charge.
Darker Weyl points correspond to greater $k_3$ values.
The darkest points correspond to the parameters of Fig.~\ref{fig:systemA_figures}b.

(iii) \emph{Creation and annihilation of Weyl points.}
Consider the scenario when the third spring is taken out of the system, which corresponds to a control vector $\mathbf{t}$ with  $k_3=0$.
In this case, there are no Weyl points in the configuration space, because the 
longitudinal normal mode has a higher frequency than the transversal mode. 
Continuously increasing the spring constant $k_3$ from zero to $k_3 = 0.76$, Weyl points are still absent. 
Increasing $k_3$ further, to $k_3 = 0.78$, we observe the creation of two Weyl points, depicted as the faintest red and blue points in Fig.~\ref{fig:systemA_figures}c.
These Weyl points move away from each other by further increasing $k_3$, as shown in Fig.~\ref{fig:systemA_figures}c.
This shows that transition patterns between different Weyl-point configurations can be studied in such a mechanical system.

(iv) \emph{Chirality flip.}
The chirality flip effect has been theoretically described in \cite{Konye}. 
The crystal studied there has a high-symmetry plane, which imposes symmetry constraints on the Weyl points and their motion as a varying mechanical strain is applied to the crystal (see Fig.~3d in \cite{Konye}).
Strain plays the role of a control parameter, and the Brillouin zone is the configuration space. 
Before applying strain, a negatively charged Weyl point resides in the high-symmetry plane of the Brillouin zone, and a mirror-symmetric pair of positively charged Weyl points resides on the two sides of the plane. 
As strain is increased, the off-plane Weyl points approach the in-plane Weyl point.
At a critical value of the strain, the three Weyl points merge, and for further increase of the strain, only a single positively charged Weyl point remains in the plane. 
From the viewpoint of the in-plane Weyl point, it has undergone a flip of its topological charge from negative to positive, hence the name `chirality flip'. 

An analogous effect is observed in System A, if we consider a special symmetric configuration of the latter, when the control parameters fulfill $r_1 = r_2$, $k_1 = k_2$, $l_1 = l_2$.
In this case, the diagonal line $\alpha = \beta$ of the configurational space is analogous to the high-symmetry plane of the Brillouin zone in \cite{Konye}.
Furthermore, the Pauli coefficients of the dynamical matrix, defined in Eq.\eqref{eq:pauli_decomposition}, have the following symmetry relations: 
\bnsn
\label{eq:symmetry}
\bean
d_z(\alpha, \beta) = d_z(\beta, \alpha) ,
\eean
\bean
d_x(\alpha, \beta) = - d_x(\beta, \alpha).
\eean
\edsn
These relations enforce a vanishing $d_x$ component on the symmetry line, that is, $d_x(\alpha,\alpha) = 0$. 
A further consequence of Eq.\eqref{eq:symmetry} is that the spectrum is symmetric, $\omega(\alpha,\beta) = \omega(\beta,\alpha)$ for both bands.
Furthermore, the Weyl points appear in the configurational space symmetrically, such that mirror-symmetric partners have the same charge (see Fig.~\ref{fig:systemA_figures}d).

The chirality flip effect in System A is illustrated in  Fig.~\ref{fig:systemA_figures}d.
For a symmetric control parameter set (see caption), we plot the Weyl points as the spring constant $k_3$ is increased. 
Initially, there are 4 Weyl points, 2 of them (faint blue) on the symmetry line, 2 of them forming a mirror pair off the symmetry line (faint red).
By increasing $k_3$, the two off-line red Weyl points approach the symmetry axis, and merge with a blue Weyl point, leaving behind a single red Weyl point (dark red) on the axis -- a clear manifestation of a chirality flip.

\section{
\label{sec:systemB}
Mechanical charge-2 Weyl point}
\begin{figure*}
    \centering
    \includegraphics{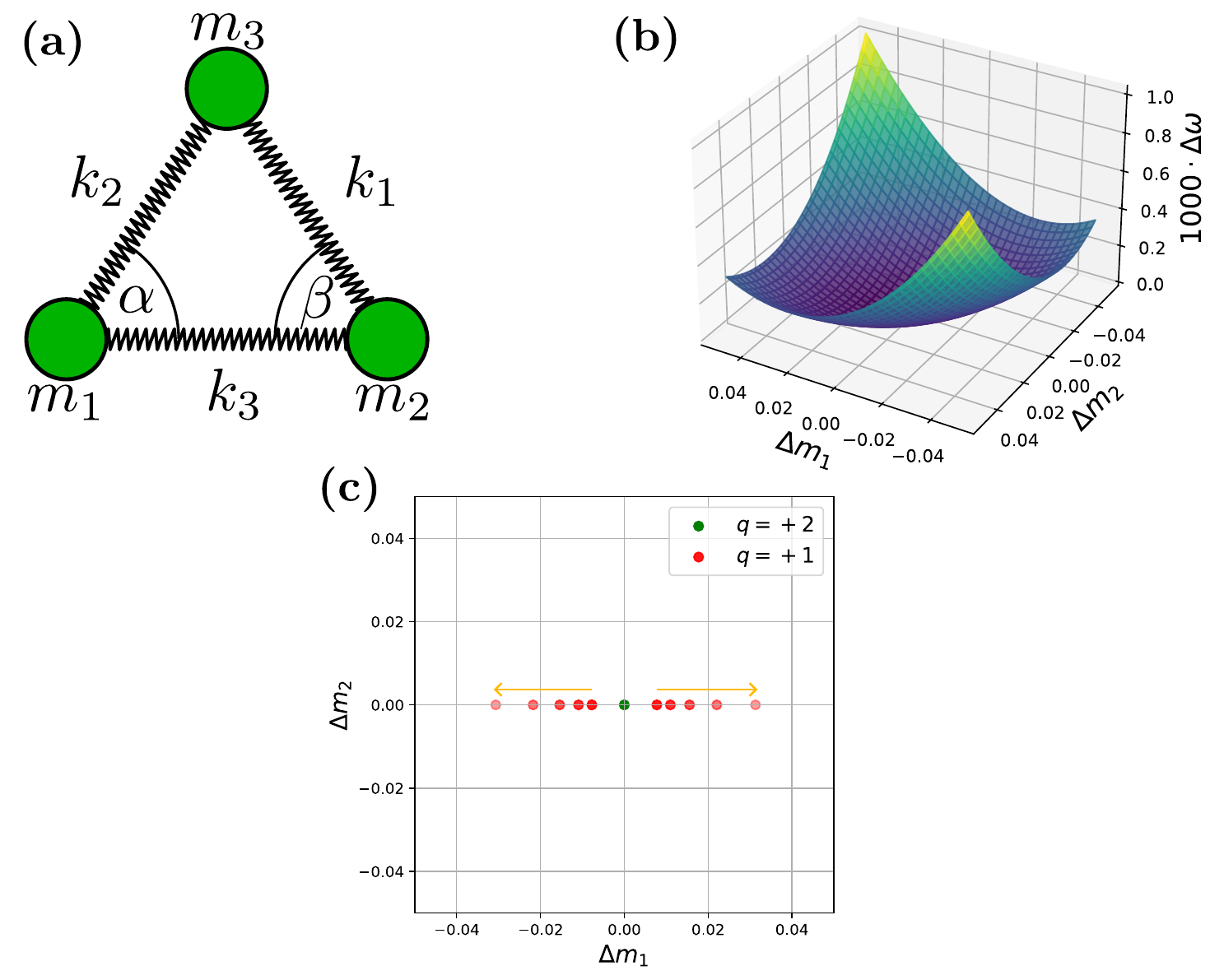}
    \caption{Quadratic degeneracy point in a mechanical system, emulating the electronic band degeneracy in bilayer graphene. 
    \textbf{(a)} Layout of System B. 
    Green circles denote masses that move in the 2D plane.
    \textbf{(b)} Frequency gap in the vibrational spectrum for System B for $k_i=1$, $m_i =1$ and $\alpha = \beta = \pi/3$, as function of the mass detunings. 
    Only the difference of the 4th and 5th eigenfrequency is shown. 
    The degeneracy splits as a quadratic function of the mass detunings $\Delta m_{1,2}$. 
    The difference of the eigenfrequencies was calculated on a $101 \times 101$ grid.
    \textbf{(c)} Evolution of Weyl points in the configuration space of mass detunings as the spring constant $\Delta k_1 = (2,4,8,16,32) \cdot 10^{-5}$ is varied.
    Control parameters: $\Delta k_2 = \Delta k_3 = 0$, $ \Delta \alpha = \Delta \beta = 0$. 
    Due to changing $\Delta k_1$, the quadratic degeneracy point dissolves into 2 Weyl points of charge $+1$, mimicking a similar effect in bilayer graphene caused by varying mechanical strain.}
    \label{fig:systemB_figures}
\end{figure*}

In the simplest tight-binding model of the electronic band structure of bilayer graphene, a non-generic degeneracy point appears at the $K$ point of the Brillouin Zone.
In the vicinity of the $K$ point, 
the electronic states are described approximately by the following effective Hamiltonian:
\cite{McCann} 
\begin{equation}
\label{eq:bilayergraphene_disprel}
    H_{\mathrm{eff}}(k_x,k_y) = \frac{\hbar^2}{2 m} \begin{pmatrix}
        0 & (k_x - i k_y)^2 \\ 
        (k_x + i k_y)^2 & 0 
    \end{pmatrix}.
\end{equation}
Here, $m$ is the effective mass of the electrons, and $k_x$ and $k_y$ are the wave vectors measured from the $K$ point. 
As discussed above, in this band structure setting, $k_x$ and $k_y$ are the configuration parameters.

In this section, we introduce a classical ball-and-spring system that emulates the non-generic degeneracy point of bilayer graphene described by Eq.~\eqref{eq:bilayergraphene_disprel}.
First, we summarize the known characteristic properties of the latter (see (i)-(iv) below), show that the ball-and-spring system indeed emulates most of those properties, and also prove a specific mathematical equivalence (linear right equivalence) between the two systems.

\subsection{Electrons in bilayer graphene}
\label{subsec:electrons_bilayergraphene}

(i) \emph{Quadratic dispersion.} 
Eq.~\eqref{eq:bilayergraphene_disprel} is an effective Hamiltonian for  electrons in bilayer graphene in the vicinity of the $K$ point. 
We will use the Pauli-matrix decomposition of this effective Hamiltonian.
This reads, omitting the constant $\hbar^2 / 2m $, as
$H_\text{eff}(k_x,k_y) = 
h_x(k_x,k_y) \sigma_x
+
h_y(k_x,k_y) \sigma_y
$,
where the coefficients are:
\bnsn
\label{eqs:definition_h_mapping}
\bean
h_x(k_y,k_y) = k_x^2 - k_y^2,
\eean
\bean
h_y(k_x,k_y) = 2 k_x k_y.
\eean
\edsn

The difference between the eigenvalues of $H_\mathrm{eff}$ is proportional to 
$ \sqrt{h_x^2 + h_y^2} = 
k_x^2 + k_y^2.
$
Hence, we say that the degeneracy at $(k_x,k_y) = 0$ splits quadratically as the function of the configurational parameters (i.e., the wave vector).

(ii) \emph{Topological charge is 2.} 
Similarly to the topological charge of the $(d_x,d_z)$ vector field, described in Sec.~\ref{sec:systemA} as a winding number around the degeneracy point in the origin of the configurational space, a topological charge is also associated to degeneracy point of the $(h_x, h_y)$ vector field. 
In fact, the topological charge of the latter is $2$. 
(Note that in a mathematical context, the term `local degree' is used for the topological charge.)

(iii) \emph{Local multiplicity is 4.}
In \cite{Pinter} it is shown that isolated twofold degeneracy points have, besides the topological charge, another characteristic, the local multiplicity.
The local multiplicity associated to such a degeneracy point is a positive integer.
In particular, for bilayer graphene, the local multiplicity associated to the vector field $(h_x,h_y)$ is 4. 

(iv) \emph{Perturbations can dissolve the quadratic degeneracy point into 2 or 4 Weyl points.}
Upon a generic `perturbation' or `deformation' of the Hamiltonian, i.e., upon a generic continuous displacement in the control space, the degeneracy point is continuously dissolved into Weyl points. 
The absolute value of the topological charge determines the minimum number of newborn Weyl points in that situation.
The local multiplicity determines the maximum number of newborn Weyl points \cite{Pinter}.
Combining these general rules with (ii) and (iii) above, it is concluded for 
the degeneracy point of bilayer graphene that perturbations can dissolve it into 2 or 4 Weyl points. 
For example, extending the simplest tight-binding model of bilayer graphene by including an additional hopping amplitude induces a perturbation to $H_\text{eff}$ which dissolves the quadratic degeneracy points into four Weyl points, known as the `trigonal warping' effect.
On the other hand, adding mechanical strain to the tight-binding model perturbs $H_\text{eff}$ such that it dissolves the quadratic degeneracy point into two Weyl points. 
Note also that the topological charge is conserved in these transitions. 

\subsection{Vibrations of System B}
\label{subsec:vibrations_systemb}

Here, we propose and study a classical ball-and-spring mechanical system whose parameter-dependent eigenfrequency spectrum shares the characteristics of bilayer graphene described in the previous section. 

The mechanical system we study here, to be called System B, is shown in Fig.~\ref{fig:systemB_figures}a. 
It consists of three point masses ($m_i$, $i \in \{1,2,3\}$), connected with three springs with spring constants $k_i$ and rest lengths $l_i$.
The masses can move only in the plane, i.e., their positions are described by 6 Cartesian coordinates altogether.
As before, we will focus on the eigenfrequencies of small, close-to-equilibrium oscillations of the system.
In equilibrium, the masses form a triangle whose geometry is determined by the rest lengths $l_i$, which are assumed to fulfill the triangle inequalities. 
Equivalently, we can characterize the triangle by two angles $\alpha$ and $\beta$, and a single rest length, i.e. $l_1$. 
We will use this latter parametrization, and will omit $l_1$, as its value does not affect the normal modes of the system.

We specify reference values for our parameters as $m_i^{(0)} =1$, $k_i^{(0)}=1$, $\alpha^{(0)}=\beta^{(0)} = \pi /3$. 
Correspondingly, we introduce \emph{detunings}, i.e.,  parameters relative to the reference values, via $\Delta k_i = k_i - k_i^{(0)}$, $\Delta m_i = m_i - m_i^{(0)}$, $\Delta \alpha = \alpha - \alpha^{(0)}$ and $\Delta \beta = \beta - \beta^{(0)}$.
These detunings can be thought of as perturbations of parameters with respect to their reference values.
We set $\Delta m_3 = 0$ and $\Delta k_3 = 0$, without the loss of generality.
Hence, the total number of parameters of the system is 6, listed as $\Delta k_{1,2}$, $\Delta m_{1,2}$, $\Delta \alpha $, and $\Delta \beta$.

We define 
the configuration space parameters as $\Delta m_{1,2}$,
and the control parameters as $\Delta k_{1,2}$, $\Delta \alpha$ and $\Delta \beta$ . 
The latter 4 parameters are collected in the control vector $\mathbf{t}$.
In our figures, we will focus on a small region of the configuration space, i.e., 
$(\Delta m_1, \Delta m_2) \in [-0.05,0.05] \times [-0.05,0.05]$.

At this point, we can anticipate the twofold spectral degeneracy of this mechanical setup, which will play the role of the twofold spectral degeneracy of $H_\mathrm{eff}(k_x =0, k_y = 0)$ of Eq.~\eqref{eq:bilayergraphene_disprel}. 
System B is described by 6 coordinates, hence its dynamical matrix, describing the normal modes, is a $6 \times 6$ real symmetric matrix, depending on the configuration and control parameters.
Consider the case when all detunings are set to zero: $\Delta m_1 = \Delta m_2 = 0$  and $\mathbf{t} = 0$.
Then the point masses form an equilateral triangle, and the symmetry group of the system is the dihedral group $D_3$. 
This group does have a two-dimensional irreducible representation (irrep), suggesting that the normal-mode eigenfrequency spectrum might have a symmetry-protected twofold degeneracy.

We do find that this is indeed the case, for this zero-detuning case, the 4th and 5th eigenfrequencies (counting from lowest to highest) are degenerate, and the modes transform according to the two-dimensional $E$ irrep of $D_3$.
This degeneracy is split as we move away from the origin of the configuration space ($\Delta m_{1,2}=0$) and hence break the symmetry. 

The $6 \times 6$ dynamical matrix of System B has three normal modes with zero eigenfrequencies. 
These correspond to the two independent translations and the single rotation of the  system. 
The remaining three normal modes have non-zero eigenfrequencies, and for $\mathbf{t} = 0$, $\Delta m_{1,2} = 0$ there is a single degenerate pair of normal modes with finite frequency.
For a fixed value of control parameters $\mathbf{t}$, we define the mapping $\omega_\mathbf{t}: \mathcal{B}_B \rightarrow \mathbb{R}_+^2$ that assigns those eigenfrequencies to each point of the configuration space, which are degenerate in the symmetric case.
Here, $\mathcal{B}_B$ denotes the configuration space. 

(i) \emph{Quadratic dispersion.}
For zero detuning of the control parameters, the splitting of the degenerate eigenfrequencies is of second order in the configuration parameters $\Delta m_{1,2}$.
This quadratic dispersion in the configuration space is illustrated in Fig.~\ref{fig:systemB_figures}b, where the difference of the eigenfrequencies, obtained from numerical diagonalization of the dynamical matrix, is plotted for $\mathbf{t}=0$.

The quadratic splitting of the degenerate frequencies can also be proven analytically, as discussed in App.~\ref{appendix:systemb}. 
To this end, we express the effective $2 \times 2$ dynamical matrix $D_\text{eff}$ of the quasi-degenerate subspace; 
the explicit form is shown in Eq.~\eqref{eq:effectivedinamicalmatrix}. 
This matrix $D_\text{eff}$ is obtained perturbatively in the configuration parameters, at the symmetric control point $\mathbf{t} = 0$, using second-order quasi-degenerate (Schrieffer-Wolff) perturbation theory.
Neglecting the unit-matrix term in the effective dynamical matrix, we obtain 
\bean
\label{eq:effective_dynamical_matrix_maintext}
\tilde{D}_\mathrm{eff} = d_x(m_x,m_y) \sigma_x + d_z(m_x, m_y) \sigma_z,
\eean
where
\bnsn
\label{eqs:definition_d_mapping}
\bean
d_x(m_x,m_y) = -\frac{\sqrt{3}}{12} m_x^2 + \frac{\sqrt{3}}{12} m_y^2,
\eean
\bean
d_z(m_x,m_y) = \frac{1}{12} m_x^2 - \frac{1}{3} m_x m_y + \frac{1}{12} m_y^2.
\eean
\edsn
Here, we have introduced the simplified notation $m_x = \Delta m_1$ and $m_y = \Delta m_2$. 
The fact that $d_x$ and $d_z$ are second order in the configuration parameter implies that quadratic dispersion.

(ii) \emph{Topological charge is 2.} This result can be obtained from a numerical evaluation of the winding number of the vector field $(d_x,d_z)$, or can be read off from a visualisation of the vector field.

(iii) \emph{Local multiplicity is 4.} This result can be obtained using any of the methods for calculation of the local multiplicity, outlined in \cite{Pinter}. 

(iv) \emph{Deformations can dissolve the quadratic degeneracy point into 2 or 4 Weyl points.}
Upon generic deformation (i.e., change of the control vector), a non-generic degeneracy point, such as the quadratic degeneracy points studied here, is dissolved to Weyl points.
The absolute value of the topological charge (local multiplicity) of the original degeneracy point is a lower (upper) bound on the number of newborn Weyl points \cite{Pinter}. 
Because of (ii) and (iii) above, we expect that deformations of System B can dissolve the quadratic degeneracy point into two or four Weyl points. 

We do confirm the two-Weyl-point scenario, which is illustrated in Fig.~\ref{fig:systemB_figures}c. 
There, the green circle depicts the quadratic degeneracy point at $\mathbf{t} = 0$.
The red points show how that is dissolved to two Weyl points, one on the left, one on the right, each with unit positive charge, as the control parameter $\Delta k_1$ is increased from zero. 

We leave it as an interesting open question if the four-Weyl-point scenario, emulating the trigonal warping effect in the bilayer graphene band structure \cite{McCann}, can be realized in this physical setting.
Without any detail, we do confirm  that the four-Weyl-point scenario can be realized in a mathematical sense by the deformation
\bean
\tilde{D}_\text{eff} \mapsto \tilde{D}_\text{eff} + \lambda ( u m_x + v m_y) \sigma_x - \lambda ( w m_x + z m_y) \sigma_z,
\eean
where the coefficients are defined in Eq.~\eqref{eqs:rightequivalence_coeffs}.
For example, we have checked numerically (not shown) that four Weyl points are born from the quadratic degeneracy point if the above deformation is applied such that $0 < \lambda < 10^{-2}$.
However, we do not know if such a deformation can be realized in a physical sense, that is, by tuning the physical parameters in the control vector $\mathbf{t}$ as
$\tilde{D}_\text{eff} \mapsto \tilde{D}_{\mathbf{t},\text{eff}}$.

\subsection{Bilayer graphene and System B are linear right equivalent}
\label{subsec:linright_equivalence}

We have just shown that the characteristic properties (i)-(iv) of the bilayer graphene effective Hamiltonian $H_\text{eff}$ are shared by those of the effective dynamical matrix $D_\text{eff}$ of System B.
Here, we show that in fact, $H_\text{eff}$
and $D_\text{eff}$ are equivalent in a specific mathematical sense: they are linear right equivalent. 
This explains the strong similarities in their properties.

The mappings $d$ and $h$ are said to be right-equivalent if there exists a diffeomorphism $f : \mathcal{B}_B \rightarrow \mathcal{B_H}$ such that the equality
\begin{equation}
\label{eq:defining_rightequivalence}
    d(m_x,m_y) = (h \circ f)(m_x, m_y)
\end{equation}
hold for all $(m_x, m_y) \in \mathcal{B}_B$.
To obtain the diffeomorphism $f$ we assume that it is a linear map and can be characterized by a $2 \times 2$ real matrix 
\begin{equation}
    F = 
    \begin{pmatrix}
        u & v \\ w & z
    \end{pmatrix},
\end{equation}
such that
\begin{equation}
\label{eq:definition_f_mapping}
    f (m_x,m_y) = F  \begin{pmatrix}
        m_x \\ m_y
    \end{pmatrix}.
\end{equation}

Inserting the defining equations of the mappings $h$ (Eq.\eqref{eqs:definition_h_mapping}), $d$ (Eq.~\eqref{eqs:definition_d_mapping}) and $f$ (Eq.~\eqref{eq:definition_f_mapping}) into Eq.~\eqref{eq:defining_rightequivalence}, and using the identifications $m_x \equiv k_x$ and $m_y \equiv k_y$, we find the following 6 equations for the unknown matrix elements of $F$:
\bnsn
\bean
u^2 - w^2 = - \frac{\sqrt{3}}{12},
\eean
\bean
v^2 - z^2 = \frac{\sqrt{3}}{12},
\eean
\bean
uv - wz = 0,
\eean
\bean
uw = \frac{1}{24},
\eean
\bean
vz = \frac{1}{24},
\eean
\bean
uz+vw = - \frac{1}{6}.
\eean
\edsn

Remarkably, the above system of equations is solvable, 
Solving the above system of equations yields
\bnsn
\label{eqs:rightequivalence_coeffs}
\bean
u = -\frac{1}{2} \sqrt{\frac{2 - \sqrt{3}}{6}},
\eean
\bean
v = + \frac{1}{2} \sqrt{\frac{2 + \sqrt{3}}{6}},
\eean
\bean
w = -\frac{1}{2} \sqrt{\frac{2 + \sqrt{3}}{6}},
\eean
\bean
z = +\frac{1}{2} \sqrt{\frac{2 - \sqrt{3}}{6}}.
\eean
\edsn
Note that the matrix $F$ defined by this solution is invertible, implying that the corresponding map $f$ is indeed a diffeomorphism. 
Note also that simultaneous sign flip of the matrix elements in Eq.~\eqref{eqs:rightequivalence_coeffs} yields an alternative solution.

With this, we have shown that the local vector fields $d$ and $h$ which describe fundamentally different physical systems are right-equivalent.
This explains the similarities of their characteristics discussed above. 

\section{
\label{sec:Discussion}
Discussion}

\subsection{Relation to prior work}

\emph{Translation invariance vs. spatial compactness.}
Engineered, macroscopic mechanical systems have already been studied to investigate topological effects arising in band structures. 
The studies we are aware of rely on the concept of translationally invariant, crystal-like metamaterials, where concepts such as wave vectors and Brillouin zone arise naturally. 
These systems involve an extensive number of degrees of freedom.
\cite{Li,Yang, Huber}
In contrast, in this work, we propose and study spatially compact mechanical setups, consisting only of a few ingredients. 
In the systems we consider, the configurational parameters are only \emph{analogous} to the wave vector. 
An inherent advantage of these setups is that only a few system elements and a few degrees of freedom have to be controlled. 

\emph{Simplification and `de-quantization' of the Weyl Josephson Circuit idea.}
This work is partly inspired by the prior proposals of emulating band-structure effects using multiply connected superconducting devices \cite{Zur_rev,Riwar,Fatemi}.
We think that spatially compact mechanical setups such as those studied in this work can provide a simplified, cost-efficient, and `de-quantized' alternative platform for such emulator experiments.
With table-top mechanical setups, the need for highly specialized fabrication and refrigeration technology is alleviated. 
Furthermore, measurement technology based on complex microwave sources and detectors for superconducting circuits can probably be substituted by more basic equipment (e.g., cameras or microphones for data acquisition, sound or ultrasound generators as driving sources), for mechanical experiments.

\emph{Exploring degeneracy points in various matrix spaces.}
Mechanical systems can also be regarded as complementary to superconducting devices, in the sense that they cover different matrix spaces. 
Namely, superconducting circuits provide access to degeneracy structures of particle-hole-symmetric \cite{Riwar} and Hermitian \cite{Fatemi} matrices, whereas spatially compact mechanical systems are described by real symmetric matrices.

\subsection{Open problems}

\emph{In-situ control of parameters in a mechanical setup.}
In our work, we study how the eigenfrequencies of coupled mechanical oscillators change as parameters are varied. 
A brute-force experimental realisation of the effects discussed here should be possible by fabricating and measuring many different samples, which have fixed but different parameter values. 
An interesting experimental challenge is to find means for in-situ parameter control.
This would alleviate the need to fabricate as many samples as many parameter settings are to be investigated. 

\emph{Frequency degeneracy points in spatially compact classical ac electronic circuits.}
Besides the mechanical setups considered here, another class of classical systems where the physics of degeneracy points can be studied is that of ac electronic circuits. 
Emulating topological materials using translational invariant circuits (`topoelectrical circuits') is a field that already exists 
\cite{Lee2018, Rafi2020}.

\emph{Trigonal warping of bilayer graphene.}
As discussed in Sec.~\ref{subsec:vibrations_systemb}, our System B emulates the quadratic degeneracy point of bilayer graphene, but we have not found a perturbation respecting the physical constraints of System B that emulates the trigonal warping effect known bilayer graphene.
This remains an interesting open problem. 

\emph{A systematic construction of mechanical systems emulating band-structure effects.}
In this work, we have identified two mechanical setups where interesting Weyl-point properties known from condensed-matter theory can be emulated. 
These mechanical setups were found \emph{intuitively}.
A natural follow-up open problem is as follows: given a condensed-matter band structure model (e.g., electronic, phononic, magnonic) with an interesting band degeneracy pattern (e.g., non-generic degeneracy point, nodal loop, nodal surface, etc.), is it possible to \emph{systematically} construct a spatially compact mechanical emulator reproducing that pattern? 
System A in our work illustrates that the torus topology of the Brillouin zone can be emulated, e.g., using suspension loops.

\emph{Replacing a cold-atom experiment with classical mechanics.}
In a recent breakthrough study \cite{Brown_2022}, the authors performed an  experiment using ultracold atoms, which used a dynamical method to probe the winding numbers of a linear and a quadratic degeneracy point in the momentum space of a honeycomb lattice. 
Such an experiment requires a highly coherent atomic ensemble and advanced control and measurement technology.
Our present work proves that degeneracy points with quadratic splitting can be engineered in simple mechanical systems, hence it highlights the opportunity of repeating the cold-atom experiment using only classical mechanics. 
Such a mechanical experiment would rely on the in-situ time-dependent tunability of the system parameters, as discussed above. 

\emph{Expanding the classification of isolated twofold degeneracy points in crystals.}
In \cite{Yu_2022}, band degeneracies in time-reversal invariant crystalline band structures were classified, and four distinct types of twofold degenerate isolated degeneracy points were identified.
In our work, we have illustrated two of those four types: Weyl points and quadratic degeneracy points (identified as \emph{charge-2 Weyl points} in \cite{Yu_2022}).
An open challenge is to engineer mechanical systems where the other two types of degeneracy points (charge-3 and charge-4 Weyl points) arise.
A further idea is to exploit the fact that spatially compact mechanical oscillators are free of the strong constraints imposed by crystal symmetries, hence they could be used to realize more `exotic' degeneracy point types which are impossible to realize in crystalline band structures. 
Similar questions arise in the context of higher-order degeneracy points, i.e., degeneracies where more than 2 normal modes share the same eigenfrequency \cite{Bradlyn}.

\section{
\label{sec:conc}
Conclusion}

We have proposed simple ball-and-spring setups which illustrate that Weyl points and their associated features, characteristic of crystalline band structures, can be emulated in classical mechanical systems.
We have shown that the parameter-dependent eigenfrequency spectrum of spatially compact ball-and-spring systems can exhibit (i) the appearance of Weyl points, (ii) the movement of Weyl points, (iii) the creation/annihilation of Weyl points, (iv) the chirality flip effect, an example of symmetry-constrained creation/annihilation, (v) quadratic degeneracy points and their dissolution to Weyl points. 
Our work opens a route toward table-top experiments on Weyl point physics, enabling the exploration of effects that have been proposed or realized with coherent quantum systems. 

\begin{acknowledgments}
This research was supported by the Ministry of Culture and Innovation and the National Research, Development and Innovation Office (NKFIH) within the Quantum Information National Laboratory of Hungary (Grant No. 2022-2.1.1-NL-2022-00004), and by the NKIFH within the OTKA Grant FK 132146.
\end{acknowledgments}

\appendix

\section{
\label{appendix:systema}
The spectrum of System A}
In this section, we discuss the calculation of the spectrum of System A. 
The calculation involves the determination of the equilibrium position of the mass, which we carry out numerically.
Hence, the dynamical matrix and the eigenfrequency spectrum is also computed numerically. 

The spectrum is defined as a map that assigns the two eigenfrequencies of the system to each angle pair $(\alpha, \beta)\in \mathcal{B}_A$.
An important step toward calculating the frequencies of small oscillations around equilibrium is to Taylor-expand
the position-dependent elastic potential around the equilibrium position. 
The elastic potential is the sum of the spring potentials:
\bean
U_\mathbf{t}(x,y) = 
\frac{1}{2}
\sum_{i=1}^3  k_i \Delta l_i ^2,
\eean
where $\Delta l_i$ is the elongation of the $i$-th spring. 
Recall that the vector $\mathbf{t}$ contains the control parameters of the system, i.e., all parameters except the angles $\alpha$ and $\beta$.
As a function of the position of the point mass, the elongations can be written as 
\bean
\Delta l_i = \sqrt{(x_i - x)^2 + (y_i-y)^2} -l_i,
\eean
where $(x_i,y_i)$ is the suspension point of the $i$-th spring, and $l_i$ is the rest length of the spring. 

The coordinates $(x_i,y_i)$ for the three springs can be written as 
\bnsn
\label{eq:suspension_coordinates}
\bean
x_1 = -d/2 -r_1 (1-\cos(\alpha)),
\eean
\bean
y_1 = r_1 \sin(\alpha),
\eean
\bean
x_2 = d/2 +r_2 (1-\cos(\beta)),
\eean
\bean
y_2 = r_2 \sin(\beta),
\eean
\bean
x_3 = 0,
\eean
\bean
y_3 = R.
\eean
\edsn

In equilibrium, the elastic potential $U_\mathbf{t}$ is minimized over the position of the point mass. 
To determine the eigenfrequencies, we  calculate the equilibrium coordinates of the body.
This is done by solving
\bean
\label{eq:nablepotequalszero}
\nabla U_\mathbf{t} = 0,
\eean
which is a system of nonlinear equations. 
We obtain the solution numerically for a specific set of parameters, using the built-in methods of Scipy. 

Given the equilibrium position of the mass $(x_0, y_0)$, its small oscillations are governed by the linearized Newton equations 
\bnsn
\label{eq:newton}
\bean
m \ddot{x} = - (\partial^2_x U_\mathbf{t}) x -  (\partial_x \partial_y  U_\mathbf{t})  y,
\eean
\bean
m \ddot{y} = -  (\partial_x \partial_y  U_\mathbf{t}) x -  (\partial^2_y U_\mathbf{t})_y,
\eean
\edsn
where the coordinates $x$ and $y$ are relative coordinates with respect to the equilibrium coordinates, and the restoring force has been linearized in the relative coordinates. 
Furthermore, in Eq.~\eqref{eq:newton}, the second-order partial derivatives of the potential $U_\mathbf{t}$ are evaluated at the equilibrium position $(x_0,y_0)$.

We collect the displacements $x$ and $y$ into a single vector \bean
\mathbf{Y} = \sqrt{m} \begin{pmatrix}
    x \\ y 
\end{pmatrix}
\eean
with which the linearized Newton equations can be written in a compact form
\bean
\mathbf{\ddot{Y}} = - \frac{1}{m} \mathcal{D}_\mathbf{t} \mathbf{Y},
\eean
where we have introduced the Hessian of the elastic potential at $(x_0, y_0)$:
\bean
\mathcal{D}_\mathbf{t} = \begin{pmatrix}
    \partial^2_x U_\mathbf{t} &  \partial_x \partial_y  U_\mathbf{t} \\ \partial_x \partial_y U_\mathbf{t} & 
    \partial^2_y U_\mathbf{t}
\end{pmatrix}.
\eean
Then, we make use of the fact that we are looking for vibrational modes fulfilling $\mathbf{\ddot{Y}} = - \omega^2 \mathbf{Y}$.
Hence, we obtain the linearized Newton equation to the following eigenvalue equation:
\bean
\omega^2 \mathbf{Y} = \frac{1}{m} \mathcal{D}_\mathbf{t} \mathbf{Y}.
\eean
The matrix on the right-hand side,
\bean
\label{eq:dynmat_hessian}
D_\mathbf{t} = \frac{1}{m} \mathcal{D}_\mathbf{t},
\eean
is the dynamical matrix of the system.

The mode eigenfrequencies are obtained by taking the square root of the eigenvalues of the dynamical matrix $D_\mathbf{t}$.
The square roots of the eigenvalues are positive as long as the Hessian is positive-definite, which is guaranteed in case of a stable equilibrium position. 
The potential $U_\mathbf{t}$ depends on the angles $\alpha$ and $\beta$ through the positions $x_{1,2},y_{1,2}$ of Eq.~\eqref{eq:suspension_coordinates} hence the above method provides the spectrum $\omega_\mathbf{t}: \mathcal{B}_A \rightarrow \mathbb{R}_+^2$ of System A. 

In Sec.~\ref{sec:systemA}, the topological charge of the Weyl points is introduced as the winding number of the vector field $(d_x, d_z)$, the latter being obtained from the Pauli decomposition of the $(\alpha, \beta)$-dependent dynamical matrix. 
This topological charge is defined as the integral
\bean
\label{eq:charge_integral}
q = \frac{1}{2 \pi} \int_{\mathcal{C}} \left( \tilde{\mathbf{d}}(\varphi)  \times \frac{d}{d \varphi} \tilde{\mathbf{d}}(\varphi) \right)_3 d \varphi,
\eean
where we have introduced
\bean
\tilde{\mathbf{d}} = \frac{\mathbf{d}}{|\mathbf{d}|}
\eean
with $\mathbf{d} = (d_x, d_z,0)$, and have used the 3D cross product ($\times$) and the notation $()_3$ referring to the third component of a three-component vector. 
The integration contour $\mathcal{C}$ encircles the degeneracy point (and only this degeneracy point) and is parametrized by the angle variable $\varphi \in [0,2\pi)$.

Below, we introduce the method we used to locate the Weyl points in the configuration space, and compute their topological charges, using the $(d_x, d_z)$ vector field.
The calculation of the topological charge is based on a discretization of the integral of Eq.~\eqref{eq:charge_integral} on a finite grid. 

First, we discretize the configuration space to create a uniform grid  of $(N+1) \times (N+1)$ points characterized by coordinates $(\alpha_j, \beta_k$). 
The spacing of the coordinates is $\delta \alpha = \delta \beta = \frac{2 \pi}{N}$ and the coordinates $(\alpha_j, \beta_k)$ correspond to the point $(-\pi + j \cdot \delta \alpha, -\pi + k \cdot \delta \beta)$ with $j,k \in \{0, 1, \dots N\}$.
We remark that by using this set of points, we overcount certain points of the configuration space, e.g. the $(\alpha_0, \beta_0) = (\alpha_0, \beta_n)= (\alpha_n, \beta_n) = (\alpha_n, \beta_0)$ all correspond to the same point of the configuration space. 

Then, we assign numbers to vertices, edges, and plaquettes of the grid as follows. 
We find the dynamical matrix and obtain the $(d_x^{(j,k)}, d_z^{(j,k)})$ vector field at the vertices of the grid,
where the superscript $(j,k)$ denotes the discrete coordinates $(\alpha_j, \beta_k)$.
Then, we calculate the phase of the vector 
$(d_x^{(j,k)},d_z^{(j,k)})$
as $\phi^{(j,k)} = \mathrm{atan_2}(d_z^{(j,k)},d_x^{(j,k)}) \in (-\pi, \pi]$.
We assign this phase to each vertex of the grid.
In the case of the edges, we calculate the phase difference between neighboring vertices $(j,k)$ and $(j',k')$ as 
\bean
\Phi^{(j,k) \rightarrow (j',k')} = \mathrm{arg}\left(\mathrm{exp} \left(i \phi^{(j',k')} - i \phi^{(j,k)} \right) \right),
\eean
where $(j',k')$ denotes a neighbour of $(j,k)$, i.e. $(j+1,k)$, $(j-1,k)$, $(j,k+1)$ or $(j,k-1)$. 
In such a way, we assign a phase difference to each oriented edge of the grid. 
Note that $\Phi$ depends on the orientation of the path $\Phi^{(j,k) \rightarrow (j',k')} \neq \Phi^{(j,k) \leftarrow (j',k')}$.

Finally, we assign the integer (vortex number)
\bean
\nonumber
Q^{(j,k)} &=& \frac{1}{2 \pi} \left( \Phi^{(j,k) \rightarrow (j+1,k)}  + \Phi^{(j+1,k) \rightarrow (j+1,k+1)} + \right. \\ 
&+& \left. \Phi^{(j+1,k+1) \rightarrow (j,k+1)} + \Phi^{(j,k+1) \rightarrow (j,k)} \right)
\eean
to the plaquette $(j,k)$.
$Q^{(j,k)}$ indicates the winding of the $(d_x,d_z)$ vector field on the boundary of the plaquette, hence it is a good indicator of the position and the topological charge of Weyl points. 
We use this technique to find Weyl points in the configuration space of System A. 

In the main text, we have analyzed the chirality flip effect, which has been  recently predicted in the context of electronic band structure theory, and which is an effect promoted by certain symmetries of the crystal. 
In the rest of the section, we discuss the symmetry properties of System A, which promote the chirality flip effect illustrated in Fig.~\ref{fig:systemA_figures}d.

Let us consider a control vector $\mathbf{t}$ such that the system has mirror symmetry upon the reflection along the $y$ axis for any $\alpha = \beta$. 
For such a mirror-symmetric setting, the elastic potential has the symmetry 
\bean
\label{eq:potential_mirrorsymmetry}
U_\mathbf{t}(x,y;\alpha, \beta) = U_\mathbf{t}(-x,y;\beta, \alpha).
\eean
Denote the equilibrium position of the mass for angle values $(\alpha,\beta)$ as $(x_0^{\alpha \beta},y_0^{\alpha \beta})$.
Then, it holds that $(x_0^{\beta \alpha},y_0^{\beta \alpha})=(-x_0^{\alpha \beta},y_0^{\alpha \beta})$. 
Equation \eqref{eq:potential_mirrorsymmetry} creates a relation between the partial derivatives of the potential at $(\alpha,\beta)$ and $(\beta, \alpha)$
\bean
\nonumber
&&\partial_x U_\mathbf{t}(\beta, \alpha) \Bigr|_{(x_0^{\beta \alpha},y_0^{\beta \alpha})} = \\ 
\nonumber
&=&\lim_{h \rightarrow 0} \frac{ U_\mathbf{t}(x_0^{\beta \alpha}+h,y_0^{\beta \alpha};\beta, \alpha)- U_\mathbf{t}(x_0^{\beta \alpha},y_0^{\beta \alpha};\beta, \alpha) }{h} = \\ 
\nonumber
&=&\lim_{h \rightarrow 0} \frac{ U_\mathbf{t}(-x_0^{\beta \alpha}-h,y_0^{\beta \alpha};\alpha, \beta)- U_\mathbf{t}(-x_0^{\beta \alpha},y_0^{\beta \alpha};\alpha, \beta) }{h} = \\ \nonumber
&=&\lim_{h \rightarrow 0} \frac{ U_\mathbf{t}(x_0^{\alpha \beta}-h,y_0^{\alpha \beta};\alpha, \beta)- U_\mathbf{t}(x_0^{\alpha \beta},y_0^{\alpha \beta};\alpha, \beta) }{h} = \\
&=& - \partial_x U_\mathbf{t}(\alpha, \beta) \Bigr|_{(x_0^{\alpha \beta},y_0^{\alpha \beta})}.
\eean

Similarly, it can be shown that the sign of the partial derivative of the potential with respect to $y$ does not change upon the interchange of the angles.
From this, we conclude that only the mixed second-order partial derivative $\partial_x \partial_y U_{\mathbf{t}}$ changes sign upon the exchange on the angles.
These relations imply that the dynamical matrix $D_\mathbf{t}(\beta, \alpha)$ is related to $D_\mathbf{t}(\alpha, \beta)$ as shown in Eq.~\eqref{eq:symmetry}.
This means that the diagonal entries are identical, while the sign of the off-diagonal entries is flipped. 
As a consequence, the eigenvalues of the 
$D_\mathbf{t}(\beta, \alpha)$ and $D_\mathbf{t}(\alpha, \beta)$
are the same, hence the spectrum is symmetric to the $\alpha = \beta$ line.
Due to the sign change of the off-diagonal entry of the dynamical matrix upon interchanging the angles, the topological charges of the symmetry-related Weyl points are the same. 
Recall that the above is valid only in case of a mirror-symmetric control parameter $\mathbf{t}$. If the mirror symmetry is broken, then the spectrum is not symmetric anymore. 

\section{
\label{appendix:systemb}
The spectrum of System B}
In this section, we discuss the calculation of the eigenfrequency spectrum of the small oscillations in System B.

The three point masses of System B are characterized by 6 coordinates.
We collect the displacements of the masses with respect to their equilibrium coordinates into a single vector $\mathbf{X} = (x_1, y_1, x_2, y_2, x_3, y_3 )^T $.
Furthermore, we define the three-component vector $\mathbf{S} = \left(S_1, S_2, S_3 \right)^T$ that contains the spring elongations, with $S_i$ being positive if the $i$-th spring is stretched and negative if it is compressed.

In the linear approximation, the spring elongations are linear functions of the displacements, i.e., $\mathbf{S} = R \mathbf{X}$, where 
$R$ is a real matrix of size $3 \times 6$, to be specified below. 
Finally, we define the vector of spring forces $\mathbf{F} = \left( F_1, F_2, F_3 \right)$. 
The component $F_i$ denotes the force exerted on the masses by the $i$-th spring.
Each force $F_i$ is regarded as a real scalar since the spring force vector is parallel to the spring itself. 
We use the convention that $F_i$ is positive when the spring is compressed.

The above definitions imply that $\mathbf{F}$ can be expressed as $\mathbf{F} = - K \mathbf{S}$, where $K$ is a real matrix of size $3 \times 3$ that contains the spring constants
\begin{equation}
    K = \begin{pmatrix}
    k_1 & 0 & 0 \\
    0 & k_2 & 0 \\ 
    0 & 0 & k_3 
    \end{pmatrix}.
\end{equation}
The spring forces couple to the displacement vector $\mathbf{X}$ via the matrix $R^\text{T}$ \cite{Rocklin}. 
Then the Newton equations of motion can be written as
\begin{equation}
\label{eq:newtonforsystemb}
    M \ddot{\mathbf{X}} = R^\text{T} \mathbf{F} = - R^\text{T} K \mathbf{S} = - R^\text{T} K R \mathbf{X}.
\end{equation}
Here $M$ is the
mass matrix of the system, namely: 
\begin{equation}
    M = \begin{pmatrix}
    m_1 & 0 & 0 & 0 & 0 & 0 \\
    0 & m_1 & 0 & 0 & 0 & 0 \\
    0 & 0 & m_2 & 0 & 0 & 0 \\
    0 & 0 & 0 & m_2 & 0 & 0 \\
    0 & 0 & 0 & 0 & m_3 & 0 \\
    0 & 0 & 0 & 0 & 0 & m_3 \\
    \end{pmatrix}.
\end{equation}

As the next step, we make use of the fact that we are looking for vibrational modes fulfilling $\ddot{\mathbf{X}} = - \omega^2 \mathbf{X}$.
Furthermore, we multiply both sides of Eq.~\eqref{eq:newtonforsystemb} by $M^{-1/2}$ from the left, to obtain the eigenvalue equation
\begin{equation}
\label{eq:dynmatrixdef}
    \omega^2 \mathbf{Y} = M^{-1/2} R^\text{T} K R  M^{-1/2} \mathbf{Y},
\end{equation}
where we have introduced the mass normalized eigenvectors $\mathbf{Y} =  M^{1/2} \mathbf{X}$.
The matrix on the right-hand side of Eq.~\ref{eq:dynmatrixdef} is the dynamical matrix $D$ of the system. 
The mass-normalized eigenvectors enforce the dynamical matrix to be symmetric.
Note that up to this point, our derivation is general, i.e., it does not exploit any symmetry assumptions for the system. 

To solve the above eigenvalue problem we need to determine the matrix $R$ which couples the displacements to the spring elongations. 
Using simple trigonometric identities we obtain
\begin{equation}
    R = \begin{pmatrix} 
    0 & 0 & \cos\beta& - \sin\beta & -\cos\beta & \sin\beta \\
    - \cos\alpha & -\sin\alpha & 0 & 0 & \cos\alpha & \sin\alpha \\ 
    -1 & 0 & 1 & 0 & 0 & 0 
    \end{pmatrix}.
\end{equation}
We emphasize that we use the notation shown in Fig.~\ref{fig:systemB_figures}a.
Using the above form of the matrix $R$ one can evaluate the matrix product in Eq.~\eqref{eq:dynmatrixdef} to obtain the dynamical matrix of the system.
The eigenvalues can be calculated for arbitrary parameter values by numerical diagonalization of the dynamical matrix.
The eigenfrequencies of the system are the square roots of the eigenvalues.

\begin{figure}
    \centering
    \includegraphics{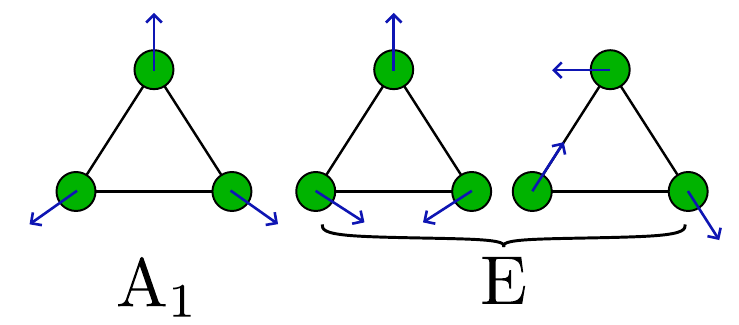}
    \caption{Normal modes of System B with non-zero eigenfrequency. The leftmost normal mode belongs to the fully symmetric irreducible representation, while the other two normal modes belong to the $E$ irrep, which is two-dimensional. These are differentiated by the eigenvalue of the vertical mirroring operation. These normal modes have $\pm 1$ eigenvalue with respect to this mirroring with the $-1$ eigenvalue corresponding to the rightmost normal mode.}
    \label{fig:normalmodes_systemB}
\end{figure}

If all the masses, springs, and angles are identical, then the symmetry group of the system is $D_3$ \cite{Tinkham2003Group}. 
Since this group does have a 2-dimensional irrep, the spectrum of System B may have a 2-fold eigenfrequency degeneracy -- and indeed, this is the case. 

In terms of irreps, one vibrational mode belongs to the irrep $A_1$, which implies that this mode is symmetric under all symmetry operations of the symmetry group. This is the so-called \textit{breathing} mode and it has the highest vibrational frequency $\omega = \sqrt{\frac{3 k }{m}}$. 
We used the simplified notation $k$ and $m$ because here, we consider the case when all the springs and masses are identical.

The other two vibrational modes belong to the $E$ irrep, which is two-dimensional, meaning that these modes are degenerate.
Their common eigenfrequency is $\omega =  \sqrt{\frac{3 k }{2 m}}$. 
 These normal modes are shown in Fig.~\ref{fig:normalmodes_systemB}.
 
The remaining 3 normal modes have zero eigenfrequency. 
These normal modes correspond (i) to the translation of the whole system along the $x$ and $y$ directions and (ii) to the rotation of the whole system along its center of mass. 
In terms of irreps, the two normal modes (i) transform between each other under the operations of $D_3$ hence correspond to the $E$ irrep, while the single normal mode (ii) corresponds to the $A_2$ irrep as the displacements change sign upon the reflection. 

We are interested in how the finite-frequency degeneracy of the modes of the $E$ irrep (shown in Fig.~\ref{fig:normalmodes_systemB}) splits as the parameters of the system are changed.
To describe this splitting, we utilize quasi-degenerate perturbation theory (Schrieffer-Wolff transformation) \cite{Winkler2003} to derive an effective $2\times 2$ dynamical matrix in the degenerate subspace.  

Due to the symmetry of the setup, we use the simplified notation $m_1^{(0)} = m_2^{(0)} = m_3^{(0)} \equiv m$ and $k_1^{(0)} = k_2^{(0)} = k_3^{(0)} \equiv k$, and fix $\alpha = \beta = \pi / 3$. 
Now we introduce the configuration-space detunings (mass detunings) $\Delta m_1$ and $\Delta m_2$, which split the twofold degeneracy.
Note that the dynamical matrix, defined via Eq.~\eqref{eq:dynmatrixdef}, depends on the masses only via the mass matrix $M$, and 
the mass detunings change only the first 4 diagonal elements in $M$. 

The diagonal entries of the matrix $M^{-1/2}$ are of the form $1/\sqrt{m + \Delta m_i}$ for $i \in \{1,2\}$ (while $\Delta m_3 = 0$). 
Then, we Taylor expand these entries up to the second order as 
\bean
\label{eq:mass_expansion}
\frac{1}{\sqrt{m + \Delta m_i}} \approx \frac{1}{m^{1/2}} - \frac{1}{2} \frac{\Delta m_i}{m^{3/2}} + \frac{3}{8} \frac{\Delta m_i ^2}{m^{5/2}}.
\eean
We do this because we anticipate that the frequency splitting is second order in $\Delta m_{1,2}$. 
The dynamical matrix can therefore be approximated as 
\bean
\label{eq:approxdynamicaldef}
D &\approx& \left(M_0^{-1/2} + M_1^{-1/2} + M_2^{-1/2 } \right) R^\text{T} K R \times \nonumber \\ 
&& \times \left(M_0^{-1/2} + M_1^{-1/2} + M_2^{-1/2 } \right),
\eean
where $M_l^{-1/2}$ contains the terms which are proportional to $\Delta m_{1,2}^l$. 
For example, $M_1^{-1/2}$ is the $6 \times 6$ matrix that contains the terms in Eq.~\eqref{eq:mass_expansion} that are first order in $\Delta m_{1,2}$, that is, 
\bean
    M_1^{-1/2} = - \frac{1}{2 m^{3/2}} \begin{pmatrix}
    \Delta m_1 & 0 & 0 & 0 & 0 & 0 \\
    0 & \Delta m_1 & 0 & 0 & 0 & 0 \\
    0 & 0 & \Delta m_2 & 0 & 0 & 0 \\
    0 & 0 & 0 & \Delta m_2 & 0 & 0 \\
    0 & 0 & 0 & 0 & 0 & 0 \\
    0 & 0 & 0 & 0 & 0 & 0 \\
    \end{pmatrix}.
\eean

As the next step, we expand the brackets in Eq.~\eqref{eq:approxdynamicaldef}, and keep terms only up to the second order in the mass detunings. 
We write the approximate dynamical matrix as 
\bean
D \approx D^0 + D^\mathrm{pert}, 
\eean
with 
\bean
D^0  = M_0^{-1/2} R^\text{T} K R  M_0^{-1/2}
\eean
and 
\bean
\nonumber
D^\mathrm{pert} &=& M_1^{-1/2} R^\text{T} K R  M_0^{-1/2} + M_0^{-1/2} R^\text{T} K R  M_1^{-1/2} + \\ 
\nonumber
&+& M_2^{-1/2} R^\text{T} K R  M_0^{-1/2} + M_1^{-1/2} R^\text{T} K R  M_1^{-1/2} + \\ 
&+& M_0^{-1/2} R^\text{T} K R  M_2^{-1/2},
\eean
where the first line contains first order terms, while the second and third line contains second order terms in $\Delta m_{1,2}$. 
We identify $D^\mathrm{pert}$ as a perturbation of $D^0$. 
This perturbation causes the degeneracy to split for nonzero $\Delta m_{1,2}$. 

To obtain an effective dynamical matrix that accounts for the splitting of the degenerate normal modes we use second-order Schrieffer-Wolff perturbation theory, which folds down the above $6 \times 6$ dynamical matrix $D$ into a $2 \times 2$ effective dynamical matrix. 
For this perturbative calculation, we need the normal modes of the symmetric system
\bnsn
\bean
\mathbf{Y}_1 = \left( 1, 0, 1, 0, 1, 0 \right)^T / \sqrt{3},
\eean
\bean
\mathbf{Y}_2 = \left( 0, 1, 0, 1, 0, 1 \right)^T / \sqrt{3},
\eean
\bean
\mathbf{Y}_3 = \left( \frac{1}{2},-\frac{\sqrt{3}}{2} , \frac{1}{2}, \frac{\sqrt{3}}{2}, -1, 0 \right)^T / \sqrt{3},
\eean
\bean
\mathbf{Y}_4 = \left( \frac{\sqrt{3}}{2}, -\frac{1}{2}, -\frac{\sqrt{3}}{2}, -\frac{1}{2}, 0, 1 \right)^T / \sqrt{3},
\eean
\bean
\mathbf{Y}_5 = \left( \frac{1}{2}, \frac{\sqrt{3}}{2}, \frac{1}{2} , - \frac{\sqrt{3}}{2}, -1, 0 \right)^T / \sqrt{3},
\eean
\bean
\mathbf{Y}_6 = \left( -\frac{\sqrt{3}}{2} ,  -\frac{1}{2}, \frac{\sqrt{3}}{2}, -\frac{1}{2}, 0, 1 \right)^T / \sqrt{3}.
\eean
\edsn

Above, $\mathbf{Y}_1$ and $\mathbf{Y}_2$ correspond to the zero frequency ($\omega_{1,2} = 0$) normal modes associated with the uniform translations along the $x$ and $y$ directions, respectively.
$\mathbf{Y}_3$ has zero eigenfrequency ($\omega_3 = 0$) as well and corresponds to the rotation of the molecule around its center of mass. 
The $\mathbf{Y}_4$ and $\mathbf{Y}_5$ normal modes belong to the $E$ irrep and are degenerate in the symmetric case with frequency $\omega_{4,5} = \sqrt{\frac{3 k}{2 m }}$.
Finally, $\mathbf{Y}_6$ is the breathing mode that belongs to the $A_1$ irrep. 
The frequency of this normal mode is $\omega_6 = \sqrt{\frac{3 k}{m}}$. 

Now we expand the $D_0$ and $D_\mathrm{pert}$ in the basis defined by the above normal modes and carry out the second order Schrieffer-Wolff transformation. 
In this basis, the matrix elements of $D_0$ and $D_\mathrm{pert}$ are written as 
\bean
D^0_{i j } = \mathbf{Y}_i^\mathrm{T} D^0 \mathbf{Y}_j
\eean
and 
\bean
D^\mathrm{pert}_{ij} = \mathbf{Y}_i^\mathrm{T} D^\mathrm{pert} \mathbf{Y}_j.
\eean
Then, the matrix elements of the effective dynamical matrix are
\bnsn
\bean
D_{\mathrm{eff}, 11} = D^0_{44} + D^\mathrm{pert}_{44} + \sum_{l = 1,2,3,6} \frac{D^\mathrm{pert}_{4l} D^\mathrm{pert}_{l4}}{\omega^2_4 - \omega^2_l},
\eean
\bean
D_{\mathrm{eff}, 12} = D^0_{45} + D^\mathrm{pert}_{45} + \sum_{l = 1,2,3,6} \frac{D^\mathrm{pert}_{4l} D^\mathrm{pert}_{l5}}{\omega^2_4 - \omega^2_l},
\eean
\bean
D_{\mathrm{eff}, 22} = D^0_{55} + D^\mathrm{pert}_{55} + \sum_{l = 1,2,3,6} \frac{D^\mathrm{pert}_{5l} D^\mathrm{pert}_{l5}}{\omega^2_4 - \omega^2_l},
\eean
\bean
D_{\mathrm{eff}, 21} = D_{\mathrm{eff}, 12},
\eean
\edsn
where we have made use of the fact that $\omega_4 = \omega_5$. 
Again, we keep terms only up to second order in $\Delta m_{1,2}$. 
With this, we obtain the effective dynamical matrix
\begin{widetext}
\bean
\label{eq:effectivedinamicalmatrix}
D_\mathrm{eff} =  \left [\frac{3 k}{2 m}  - \frac{1}{2} \frac{k}{m^2} \left( \Delta m_1 + \Delta m_2 \right) \right] \sigma_0 
+ \frac{k}{m^3}\begin{pmatrix}
\frac{5}{12} \Delta m_1^2 - \frac{1}{6} \Delta m_1 \Delta m_2 + \frac{5}{12} \Delta m_2^2 & - \frac{\sqrt{3}}{12} \Delta m_1^2 + \frac{\sqrt{3}}{12} \Delta m_2^2 \\ 
- \frac{\sqrt{3}}{12} \Delta m_1^2 + \frac{\sqrt{3}}{12} \Delta m_2^2  & \frac{1}{4} \Delta m_1^2 + \frac{1}{2} \Delta m_1 \Delta m_2 + \frac{1}{4} \Delta m_2^2
\end{pmatrix},
\eean
\end{widetext}
where $\sigma_0$ the $2 \times 2$ identity matrix.
The above matrix describes the splitting of the degenerate normal modes. 
The first term is proportional to the identity matrix and hence causes no splitting between the normal modes.
In contrast, the second term contains $\sigma_x$ and $\sigma_z$ contributions as well and depends quadratically on the configuration parameters. 
We identify the traceless part of the second term as the effective dynamical matrix $\tilde{D}_\mathrm{eff}$ of Eq.~\eqref{eq:effective_dynamical_matrix_maintext}.
This effective dynamical matrix is valid only in the vicinity of the origin. 

The symmetry of the system can be broken by changing the control parameters of the system, for example, $\Delta k_1 \neq 0$. 
In this case, the charge-2 Weyl point splits into regular Weyl points (a.k.a.~charge-1 Weyl points). 
In the main text, we have shown results related to these Weyl points. 
To find the Weyl points numerically, we use the method discussed in App.~D. of \cite{Frankteleportation}. 
Then, for each Weyl point found, we carry out a numerical Schrieffer-Wolff transformation to obtain the effective dynamical matrix around the Weyl point. 
The charge of the Weyl point is the winding of the vector field defined by the effective dynamical matrix.  
Here, it is important to keep the orientation of the quasi-degenerate subspace fixed as the control parameters are varied because the winding of the vector field does depend on the orientation of the subspace. 
This orientation can be fixed by choosing the sign of the normal modes consistently. 

\bibliography{ref}
\end{document}